\documentclass[aps,prd,preprintnumbers,showpacs,showkeys,nofootinbib,
superscriptaddress,fleqn,floatfix,tightenlines]{revtex4-1}
\usepackage{amsmath,amsfonts,amssymb,amscd,amsxtra,amsthm}
\usepackage{graphicx}  
\usepackage{epstopdf}
\usepackage{dcolumn}  
\usepackage{bm}          
\usepackage{slashed}
\usepackage[utf8]{inputenc} 

\usepackage[normalem]{ulem} 
\usepackage[dvipsnames]{xcolor} 
\renewcommand\sout{\bgroup \color{red} \ULdepth=-.5ex \ULset}

\begin{document}
\preprint{INHA-NTG-02/2016}
\title{Instanton effects on the heavy-quark static potential} 

\author{
U.T.~Yakhshiev}
\email{yakhshiev@inha.ac.kr}
\affiliation{
Department of Physics, Inha University,
Incheon 22212, Republic of Korea}
\affiliation{RIKEN Nishina Center, RIKEN, 2-1 Hirosawa, 351-0115
  Saitama, Japan} 

\author{Hyun-Chul~Kim}
\email{hchkim@inha.ac.kr}
\affiliation{
Department of Physics, Inha University,
Incheon 22212, Republic of Korea}
\affiliation{RIKEN Nishina Center, RIKEN, 2-1 Hirosawa, 351-0115
  Saitama, Japan} 
\affiliation{School of Physics, Korea Institute for Advanced Study
  (KIAS), Seoul 02455, Republic of Korea}

\author{B.~Turimov}
\email{bturimov@yahoo.com}
\affiliation{
Department of Physics, Inha University,
Incheon 22212, Republic of Korea}

\author{
M.M.~Musakhanov}
\email{yousuf@uzsci.net}
\affiliation{
Theoretical Physics Department, National University of Uzbekistan, 
Tashkent-174, Uzbekistan}

\author{Emiko Hiyama}
\email{hiyama@riken.jp}
\affiliation{RIKEN Nishina Center, RIKEN, 2-1 Hirosawa, 351-0115
  Saitama, Japan}

\date{June, 2016}

\begin{abstract}
We investigate the instanton effects on the heavy-quark potential,
including its spin-dependent part, based on the instanton liquid
model. Starting with the central potential derived from the instanton
vacuum, we obtain the spin-dependent part of the heavy-quark
potential. We discuss the results of the heavy-quark  
potential from the instanton vacuum. We finally solve the
nonrelativistic two-body problem, associating with the heavy-quark
potential from the instanton vacuum. The instanton effects on the
quarkonia spectra are marginal but are required for quantitative
description of the spectra. 
\end{abstract}


\maketitle

\section{Introduction}
Heavy-quark physics has evolved into a new phase. Charmonium-like
states, which are known as XYZ states~\cite{Choi:2003ue,
  Aubert:2004ns, Aubert:2005rm, Abe:2007jna, Choi:2007wga,
  Belle:2011aa, Liu:2013dau, 
  Ablikim:2013mio, Ablikim:2013wzq,
  Aaij:2013zoa,Ablikim:2013xfr,Aaij:2014jqa, 
  Aaij:2015zxa} and quite possibly exotic ones, conventional
bottomonia including the lowest-lying state
$\eta_b$~\cite{Aubert:2008ba, Aubert:2009as, Bonvicini:2009hs,
  Mizuk:2012pb, Dobbs:2012zn, Tamponi:2015xzb, Abdesselam:2015zza},
and heavy pentaquark states~\cite{Aaij:2015tga} have been newly
reported by various experimental collaborations (see also recent 
reviews~\cite{Bevan:2014iga, Andronic:2015wma,Yuan:2015kya,  
  Yuan:2015ztu}). These novel findings of heavy hadrons have renewed
interest in heavy-quark spectra and have triggered subsequently a
great deal of experimental and theoretical work (see for example the
following reviews~\cite{Swanson:2006st, Eichten:2007qx,
  Voloshin:2007dx, Brambilla:2010cs,Olsen:2014qna}).      
Among these newly observed heavy hadrons, the conventional 
bottomonium  $\eta_b(1S)$ is placed in a crucial position. Even though 
it is the lowest-lying bottomonium, it has been observed only very
recently~\cite{Aubert:2008ba, Aubert:2009as, Bonvicini:2009hs,
  Mizuk:2012pb, Dobbs:2012zn} and the precise measurement of its mass
provides a subtle test for any theory about heavy quarkonia, based on  
quantum chromodynamics (QCD)~\cite{Penin:2009wf, Recksiegel:2003fm,
  Kniehl:2003ap}.    

Various theoretical methods for the quarkonium spectra have been
developed over decades (see recent reviews~\cite{Eichten:2007qx,
  Voloshin:2007dx, Brambilla:2010cs, Patrignani:2012an}), 
among which the potential model has been widely used for describing 
properties of the quarkonia~\cite{Eichten:1974af, Eichten:1978tg}. The 
form of the potential at short distances is governed by the
Coulomb-like interaction arising from perturbative QCD (pQCD). In the
lowest order, one-gluon exchange between a heavy quark and a heavy
anti-quark is responsible for this Coulomb-like
attraction~\cite{Susskind:1976pi, Appelquist:1977tw, 
  Appelquist:1977es, Fischler:1977yf}. The running coupling constant
for the Coulomb-like interaction was considered with higher order
corrections in pQCD~\cite{Peter:1996ig, Peter:1997me, Schroder:1998vy,
  Smirnov:2009fh, Anzai:2009tm}. However, the distance of the quark
and the anti-quark gets farther apart, certain nonperturbative
contributions should be taken into account in the potential. Quark
confinement~\cite{Wilson:1974sk} is shown to be the most essential
nonperturbative part obtained at least phenomenologically from the
Wilson loop for the heavy-quark potential, which rises linearly at
large distances~\cite{Eichten:1974af, Eichten:1978tg}.   
This linearly rising potential was extensively studied in lattice
QCD~\cite{Bali:1992ab,Booth:1992bm, Bali:1996cj, Glassner:1996xi,
  Bali:1997am, Bali:2000gf, Kawanai:2013aca, Kawanai:2015tga}.  

There are yet another nonperturbative effects on the heavy-quark
potential from instantons~\cite{Belavin:1975fg}, which are known to be
one of the most important topological objects in describing the QCD
vacuum. These instanton effects on the heavy-quark potential were
already studied many years ago~\cite{Wilczek:1977md,Callan:1978ye, 
  Eichten:1980mw}, spin-dependent aspects of the heavy-quark potential
being emphasized. The central part of the heavy-quark potential was
first derived~\cite{Diakonov:1989un}, based on the instanton liquid
model for the QCD vacuum~\cite{Diakonov:1983hh, Diakonov:1985eg,
  Diakonov:2002fq}. In Ref.~\cite{Diakonov:1989un}, the Wilson loop
was averaged in the instanton ensemble to get the heavy-quark
potential, which rises almost linearly as the relative distance
between the quark and the antiquark increases, then it starts to get
saturated. The results of Ref.~\cite{Diakonov:1989un} were also
simulated in lattice QCD~\cite{Fukushima:1997rc, Chen:1998ct,
  Diakonov:1998rk}. Though the instanton vacuum does not explain quark
confinement,  it will play a certain role in describing the
characteristics of the quarkonia. The feature of the instanton vacuum will
be recapitulated briefly in the present work in the context of the
quarkonium hyperfine mass splittings.     

In this work, we will examine the instanton effects on the heavy-quark
potential from the instanton vacuum, including the spin-dependent
parts in addition to the central one. In fact, Eichten and 
Feinberg~\cite{Eichten:1980mw} derived an analytic form of the 
instanton contributions to the spin-dependent potential but were not
able to compute them due to difficulties of deriving the static energy
or the central static potential induced from instantons. 
Diakonov et al.~\cite{Diakonov:1989un} calculated this central part of
the heavy-quark potential from the instanton vacuum, as mentioned
previously. Thus, in the present work, we want to obtain the
instanton-induced spin-dependent parts of the heavy-quark potential,
following closely Refs.~\cite{Diakonov:1989un, Eichten:1980mw}.   
To derive the spin-dependent potential from the instanton vacuum, we
first expand the matter part of the QCD Lagrangian for the heavy quark
with respect to the inverse of a heavy-quark mass ($1/m_Q$), as
usually was done in heavy-quark effective theory (HQET). As was
obtained from Ref.~\cite{Diakonov:1989un}, the central
part comes from the leading order in the heavy-quark expansion. 
The heavy-quark propagator or the Wilson loop being averaged over the 
instanton medium, the central part can be derived. The spin-dependent
contributions arise from the order of $1/m_Q^2$. As we will show in
this work, the heavy-quark propagator is given as an integral
equation. Expanding it in powers of $1/m_Q^2$, we are able to compute
the spin-dependent part of the heavy-quark potential as was first
shown in Ref.~\cite{Eichten:1980mw}. We will evaluate these
spin-dependent potentials and examine their behaviour. Then we will
proceed to compute the instanton effects on the hyperfine mass
splittings of quarkonia. Assuming that the interaction range between a
heavy quark and a heavy anti-quark is smaller than the inter-instanton
distance, we can easily deal with the effects of the instantons on the
hyperfine mass splittings of the quarkonia. We find at least
qualitatively that the instantons have definite effects on those of
the charmonia, while those of the bottomonia acquire tiny effects from
the instanton vacuum because of the heavier mass of the bottom quark.     

The paper is organized in the following way. In the next 
section~\ref{sec:Formalism}, we explain how to derive the instanton
effects on the heavy-quark potential systematically. We first
review the results of Ref.~\cite{Diakonov:1989un} within the
heavy-quark expansion. Then we show the corrections to the
spin-dependent heavy-quark potential, which come from the
$1/m_Q^2$ order.
In Section~\ref{sect:Results} we discuss the results
of the instanton effects on the heavy-quark potential in detail and present 
numerical method used to solve the Schr\"{o}dinger equation. We
also present the spectrum low laying charmonium states and the estimates
 of the hyperfine mass splittings of these states. 
Finally, in Section~\ref{sect:Summary} we summarise
the results and give a future outlook related to the present work.  

\section{Formalism}
\label{sec:Formalism}
\subsection{Heavy-quark propagator}
We start with the matter part of the QCD Lagrangian for the heavy
quark, given as  
\begin{align}\label{FL}
{\cal L}_{\Psi} =\bar{\Psi}(x)\left(i \slashed{D} -m_Q\right) \Psi(x),
\end{align}
where $i\slashed{D}=i\slashed\partial +\slashed A$ denotes the
covariant derivative, $m_Q$ stands for the mass of the heavy quark,
and $\Psi(x)$ represents the field corresponding to the heavy
quark. As was done in HQET~\cite{Georgi:1990um, Mannel:1991mc}, we
assume that the heavy-quark mass $m_Q$ goes to infinity with the
velocity $v$ of the heavy quark fixed ($v^2=1$). Then we can decompose
the heavy-quark field into the large component $h_v(x)$ and the small
one $H_v(x)$ as follows  
\begin{align}\label{FWT}
\Psi(x)=e^{-im_Q v\cdot x}\big[h_v(x)+H_v(x)\big],
\end{align}
which is just the Foldy-Wouthuysen transformation~\cite{Foldy:1949wa,
  Korner:1991kf} used in the nonrelativistic expansion in QED.  
The $h_v(x)$ and $H_v(x)$ fields are defined respectively as  
\begin{align}
\label{Q}
h_v(x)&=e^{im_Qv\cdot x}\left(\frac{1+\slashed v}{2}\right)\Psi(x),\\
  \slashed v h_v(x) & = h_v(x), \cr
\label{h}
H_v(x)&=e^{im_Qv\cdot x}\left(\frac{1-\slashed v}{2}\right)\Psi(x), \\
  \slashed v H_v(x)& = -H_v(x).\nonumber
\end{align}
The velocity vector allows one also to split the covariant
derivative into the longitudinal and transverse components as 
\begin{align}
\slashed{D} = \slashed{v}(v\cdot D) + \slashed{D}_\perp,   
\end{align}
where $\slashed{D}_\perp = \gamma^\mu(g_{\mu\nu}  - v_\mu
v_\nu)D^\nu$. The transverse component of the covariant derivative
satisfies the relations
\begin{align}
(i\slashed{D}_\perp)^2 = -{\bm D}^2 + \frac{1}{2} \sigma \cdot G =
                         {\bm P}^2 + \bm \sigma \cdot \bm B,  \;\;\;\;
i\slashed{D}_\perp (iv\cdot D) i\slashed{D}_\perp
= \bm E \cdot \bm D + \bm \sigma \cdot(\bm E\times \bm D),
\label{eq:RelationsForD}
\end{align}
where $G_{\mu\nu}$ stands for the gluon field strength tensor. $\bm E$
and $\bm B$ denote the chromoelectric and chromomagnetic fields,
respectively. Using the equations of motion, we can remove the small
field $H_v(x)$ by the relation
\begin{align}\label{hQ}
H_v=\frac{1}{2m_Q+iv\cdot D}i\slashed D_\perp h_v
\end{align}
or equivalently we can integrate out the $H_v$
fields~\cite{Mannel:1991mc}.  Thus, we arrive at the 
effective action expressed only in terms of the $h_v$ fields 
\begin{align}
{\cal S}_{\mathrm{eff}}[h_v,\,A] = \int\limits d^4 x\,
\bar h_v\left[iv\cdot D - i\slashed D_\perp
\frac{1}{2m_Q+iv\cdot D}i\slashed D_\perp \right]h_v,
\label{LHQET}
\end{align}
where the first term will provide the central contribution to the
heavy-quark potential while the second term is responsible for the 
spin-dependent part.

Using the effective Lagrangian given in Eq.~(\ref{LHQET}), we can
define the heavy quark propagator as 
  \begin{align}
\label{Sr}
\left[iv\cdot D - i\slashed D_\perp\frac{1}
{2m_Q+iv\cdot D}\,i\slashed D_\perp\right]
 S(x,y;A)=\delta^{(4)}(x-y).    
  \end{align}
If we assume that the heavy-quark mass is infinitely heavy, then the
heavy-quark propagator in the leading order 
satisfies the following equation 
\begin{align}\label{Snr}
(iv\cdot D) S_0(x,y;A) = \delta^{(4)}(x-y)
\end{align}
and its solution in the rest frame $v=(1,\,\bm 0)$ is found to be
\begin{align}
S_0(x,y;A_4) = P
\exp{\left(i\int_{x_4}^{y_4} {\rm d}z_4 A_4 \right)} 
\delta^{(3)}(\bm x-\bm y),
\label{eq:LO_propagator}
\end{align}
where $A_4$ is the time component of the gluon field in
four-dimensional Euclidean space. Note that since we consider the
instanton field, which is the classical solution in Euclidean space,
we work in Euclidean space from now on. Equation~(\ref{eq:LO_propagator})
implies that the heavy quark propagates along the time direction. The
full propagator $S(x,y;A)$ is then expressed as an integral equation
as follows  
\begin{align}
S(x,y;A) = S_0(x,y,A)-\int {\rm d}^4z\, S_0(x,z;A)
\left[i\slashed D_\perp\frac{1}{2m_Q+iv\cdot D}i\slashed
   D_\perp\right]S(z,y;A). 
\label{eq:FullPropagator}
\end{align}
Since $m_Q$ is rather heavy, we can expand iteratively the full
propagator (\ref{eq:FullPropagator}) in powers of $1/m_Q$, when we
derive the spin-dependent heavy-quark potential.

\subsection{Heavy-quark potential from the instanton vacuum} 
\begin{figure}[h]
\vspace{0.5cm}
\includegraphics[width=9cm,angle=0]{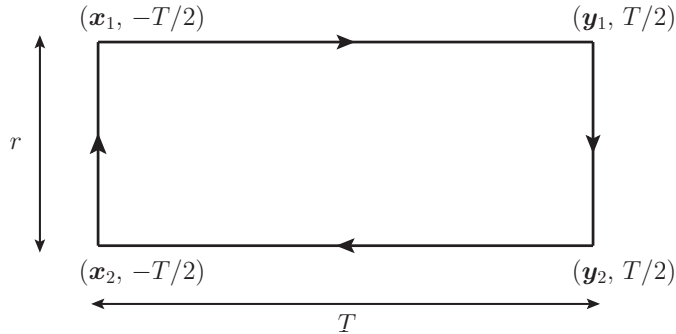}\\
\caption{The rectangular Wilson loop.}
\label{fig1}
\end{figure}

The static heavy-quark potential is defined as
the expectation value of the Wilson loop in a manifestly
gauge-invariant manner  
\begin{align}
V(r)= - \lim _{T\rightarrow \infty} \frac{1}{T} \ln\left\langle
  0\left| \mathrm{Tr}\left\{W_C[A]\right\}\right|0\right\rangle,
\label{eq:WilonPot1}
\end{align}
where $W_C[A]$ denotes the Wilson loop expressed as  
\begin{align}
W_C[A] =   P\exp{\left(i\oint_{C} {\rm d}z_{\mu}
    A_{\mu}(z) \right)}.
\end{align}
The path is usually taken to be a large rectangle ($T\times r$) as
drawn in Fig.~\ref{fig1} with $r=|\bm x_1-\bm x_2|=|\bm y_1-\bm y_2|$.   
We first consider the central potential from the instanton vacuum, 
restating briefly the results from Ref.~\cite{Diakonov:1989un}.
The leading-order expectation value of the Wilson loop in Euclidean 
space is defined as    
\begin{align}
  \label{eq:ExpectationValue}
\langle W_C[A]\rangle = \int DA_\mu  \mathrm{Tr}
  P\exp{\left(i\oint_{C} {\rm d}x_{\mu}  A_{\mu}(x) \right)}
  e^{-\mathcal{S}_{\mathrm{YM}}}
\end{align}
where $S_{\mathrm{YM}}$ is the Yang-Mills action for the gluon
field. The Wilson loop in the instanton medium can be written as 
\begin{align}
W_C[I,\bar{I}] =  P\exp{\left(i\oint_{C}   {\rm d}t \sum_{I,\bar{I}}
  a_{I,\bar{I}} \right)},   
\end{align}
where $a_{I,\bar{I}}= \dot{x}_{\mu} A_{\mu}^{I,\bar{I}}(x)$. $I$
($\bar{I})$ denotes the instanton (anti-instanton). 
$A_{\mu}^{I,\bar{I}}$ represent the instanton (anti-instanton)
solutions of which the explicit expressions can be found in 
Appendix. 
The sum $\sum_{I,\bar{I}}a_{I,\bar{I}}$ stands for the superposition
of $N_+$ instantons and $N_-$ anti-instantons for the classical gluon 
background field $\bar{A}_\mu$, which is written as  
\begin{align}
\dot{x}_{\mu} \bar{A}_\mu (x,\,\xi) =  \sum_{I=1}^{N_+} a_I (x,\,\xi) +
  \sum_{\bar{I}=1}^{N_-}  a_{\bar{I}}(x,\,\xi), 
\end{align}
where $\xi$ represents the set of collective coordinates for the
instanton, consisting of its center $z_{I\mu}$, the size $\rho_I$, and
$\mathrm{SU}(N_c)$ orientation matrix with the number of colors
$N_c$. The integration over the gluon fields given in
Eq.~(\ref{eq:ExpectationValue}) is then replaced with the integrations
over the set of collective coordinates of the instantons 
(anti-instantons)~\cite{Diakonov:1983hh, Diakonov:1985eg,
  Diakonov:1989un} such that Eq.~(\ref{eq:ExpectationValue}) can be
understood as an average over instanton ensemble.  

The leading-order heavy-quark propagator in the rest  
frame is written in terms of the superposition of the instantons   
\begin{align}
S_0^{(i)}(x,y;a_{I,\bar{I}}) = \langle y| \left(\frac{d}{dt} -
  \sum_{I,\bar{I}} a_{I,\bar{I}}^{(i)}  +i \epsilon \right)^{-1} |x\rangle,
\label{eq:InstExpand}
\end{align}
where $a_{I,\bar{I}}^{(i)}$ represents the gluon field projected onto 
the corresponding \textit{i}th Wilson line. 
Since $T\to \infty$, we can neglect the short sides of the rectangular
path. The separation between the two long Wilson lines is given 
as $r$, as shown in Fig.~\ref{fig1}. 
Using Eqs.(\ref{eq:LO_propagator}) and (\ref{eq:InstExpand}), we can
write the Wilson loop along the rectangle shown in Fig.~\ref{fig1} as  
\begin{align}
\mathrm{Tr} W_C = \left
  \langle\kern-3\nulldelimiterspace\left \langle \mathrm{Tr}\left[ 
S_0^{(1)}(\bm x_1,-T/2, \bm y_1,T/2\; ; a_{I,\bar{I}}) S_0^{(2)}(\bm
  x_2,-T/2, \bm y_2,T/2\; ;a_{I,\bar{I}})   \rangle\right] 
\right\rangle\kern-3\nulldelimiterspace\right \rangle, 
\label{eq:WilsonTwoLines}
\end{align} 
The double angle bracket $\left \langle\kern-5\nulldelimiterspace\left
    \langle \cdots   \right\rangle\kern-5\nulldelimiterspace\right
\rangle$ emphasizes the average over the instanton ensemble. Each
heavy-quark propagator in Eq.~(\ref{eq:WilsonTwoLines}) is 
expanded in powers of the instanton and anti-instanton fields
$a_{I,\bar{I}}^{(1,2)}$. Then the sum of the planar diagrams is
carried out, which is the leading order in the $1/N_c$
expansion~\cite{Pobylitsa:1989uq}. Note that the instanton vacuum has
two parameters characterizing the dilute instanton
liquid~\cite{Shuryak:1981ff, Diakonov:1983hh}: the average 
size of the instanton $\bar{\rho}\simeq 
0.33\,\mathrm{fm}$ and the average separation between instantons
$\bar{R}=(N/V)^{-1/4}\simeq 1\,\mathrm{fm}$, where the instanton
density is given as $N/V\simeq (200\,\mathrm{MeV})^{4}$. 
It allows one to use $N/VN_c$ as a small perturbation parameter. 
We refer to Ref.~\cite{Diakonov:1989un} for further details of the 
calculation.  

Using Eq.~(\ref{app:2}), we can obtain the explicit form of the central
potential from the instanton vacuum as
\begin{align}
V_C&=\frac{N}{2VN_c} \int d^3 z_I \mathrm{Tr}_c \left[1-
       P\exp\left(i\int_{-T/2}^{T/2} dx_4 A_{I4}^{(1)} \right)
       P\exp\left(-i\int_{-T/2}^{T/2}  dx_4 A_{I4}^{(2)}\right)
       \right]_{z_{I4}=0} + (I\to \bar{I}) \cr
&=\frac{2N}{VN_c}\int d^3z\left[1-\cos\left(\frac{\pi |\vec z|}{\sqrt{|\vec z|^2+\bar\rho^{\,2}}}\right)
\cos\left(\frac{\pi |\vec z+\vec r|}{\sqrt{|\vec z+\vec r|^2+\bar\rho^{\,2}}}\right)\right.\cr
&\left.-\frac{\vec z(\vec z+\vec r)}{|\vec z||\vec z+\vec r|}\sin\left(\frac{\pi |\vec z|}{\sqrt{|\vec z|^2+\bar\rho^{\,2}}}\right)
\sin\left(\frac{\pi |\vec z+\vec r|}{\sqrt{|\vec z+\vec r|^2+\bar\rho^{\,2}}}\right)
\right],
\end{align}
where $z$ denotes the position of the instanton, which
is one of the collective coordinates for the instantons. The trace
$\mathrm{Tr}_c$ runs over the colour space and
$r$ is a distance between quark and antiquark.
 Further introducing 
the dimentionless variables $y=z/\bar \rho$ and $x=r/\bar\rho$, one can rewrite 
the potential in terms of the dimensionless integral $I(x)$
\begin{eqnarray}
\label{eq:LO_pot}
V_C(r)&=&\frac{4\pi N\bar\rho^{\,3}}{VN_c}\,I\left(\frac{r}{\bar\rho}\right),\\
\label{dimint}
I(x)&=&\int_0^\infty y^2dy\int_{-1}^1dt\,
\left[1-\cos\left(\pi\frac{ y}{\sqrt{y^2+1}}\right)
\cos\left(\pi\sqrt{\frac{y^2+x^2+2xyt}{y^2+x^2+2xyt+1}}\,\right)\right.\cr
&&\left.-\frac{y+xt}{\sqrt{y^2+x^2+2xyt}}\sin\left(\pi\frac{ y}{\sqrt{y^2+1}}\right)
\sin\left(\pi\sqrt{\frac{y^2+x^2+2xyt}{y^2+x^2+2xyt+1}}\,\right)
\right].
\end{eqnarray}
 As $r$ goes to infinity, the potential is saturated to be a constant  
\begin{align}
\lim_{r\to\infty} V_C(r) = 2\Delta M_Q,
\end{align}
where $\Delta M_Q$ is the correction to the heavy-quark mass from the
instanton vacuum ~\cite{Diakonov:1989un}
\begin{align}
\Delta M_Q &= \frac{N}{2VN_c} \int d^3 z \mathrm{Tr}_c\left[ 
1- \left.P \exp\left(i\int\limits_{-\infty}^\infty dx_4
             A_{I4}\right) \right|_{z_4=0} \right] + (I\to \bar{I})
\cr
&= \frac{8 \pi N\bar{\rho}^3}{VN_c} \int\limits_0^\infty dy\,y^2
  \left(1+\cos \frac{\pi y}{\sqrt{y^2+1}}\right)=-\frac{4 \pi^4 N\bar{\rho}^3}{3VN_c}
  \left(J_0(\pi)+\frac1{\pi}J_1(\pi)\right) 
\end{align}
calculated using again Eq.~(\ref{app:2}).  The average
size of the instanton is regarded as a renormalization scale of the
instanton vacuum~\cite{Diakonov:1985eg, Son:2015bwa}. 
Keeping in mind the fact that the current quark mass is
scale-dependent and its value is usually given at $\mu=m_c$, certain
scaling effects arising from the renormalization group equation for
the quark mass should be taken into account in order to estimate the
effects on the heavy-quark mass from the instanton vacuum.  The
instanton effects would be slightly decreased, when one matches the
scale of $\Delta M_Q$ to the charmed quark mass given in
Ref.~\cite{PDG2016}.  

We are now in a position to consider the spin-dependent parts of the
heavy-quark potential. The general procedure is very similar to what
was done in Eq.~(\ref{eq:WilsonTwoLines}). Since we consider now the
finite heavy-quark mass, we need to use the full propagator given in
Eq.~(\ref{eq:FullPropagator}) instead of the leading one. That is, we
calculate the two Wilson lines as 
\begin{align}
\mathrm{Tr} W_C = \left
  \langle\kern-3\nulldelimiterspace\left \langle \mathrm{Tr}\left[ 
S^{(1)}(\bm x_2,-T/2, \bm y_2,T/2\; ; a_{I,\bar{I}}) S^{(2)}(\bm
  x_1,-T/2, \bm y_1,T/2\; ;a_{I,\bar{I}}) \rangle\right] 
\right\rangle\kern-3\nulldelimiterspace\right \rangle. 
\label{eq:WilsonFullLines}
\end{align}
Considering the fact that $1/m_Q$ can be regarded as a small
parameter, we can expand the full propagators in
Eq.~(\ref{eq:WilsonFullLines}) iteratively in powers of $1/m_Q$. 
Using the relations given in Eq.~(\ref{eq:RelationsForD}), we  
first expand the term between $S_0$ and $S$ in powers of $1/m_Q$
\begin{align}
i\slashed D_\perp\frac{1}{2m_Q+iv\cdot D}i\slashed
   D_\perp \approx \frac1{2m_Q}\left(-\bm D^2 +\sigma \cdot \bm B\right)
+ \frac1{4m_Q^2} \left[\bm E\cdot \bm D + \bm\sigma \cdot (\bm E\times
  \bm D) \right]. 
\end{align}
Then, the heavy-quark propagator for the $i$th Wilson loop can be
iteratively expressed in powers of $1/m_Q$ as  
\begin{align}
S^{(i)}(x,y;A) & \approx S_0^{(i)}(x,y;A) - \frac1{2m_Q}\int d^4 \eta\,
                 S_0^{(i)} (x,\eta;A)
 (-\bm D^2 +\bm \sigma_i \cdot \bm B) S_0^{(i)}(\eta,y;A) \cr
&-\frac1{4m_Q^2} \int d^4 \eta \,S_0^{(i)} (x,\eta;A) (\bm D\cdot \bm D +\bm
  \sigma_i \cdot (\bm E\times \bm D)) S_0^{(i)} (\eta,y;A) \cr
&+  \frac1{4m_Q^2}\int d^4 \eta d^4 \eta'\, \theta(\eta_4'-\eta_4)
  S_0^{(i)}(x,\eta;A)  (-\bm D^2 +\bm \sigma_i \cdot \bm B)
 S_0^{(i)} (\eta,\eta';A) \cr
& \hspace{3cm} \times\, (-\bm D^2 +\bm \sigma_i \cdot \bm B) S_0^{(i)}
  (\eta',y;A).  
  \label{eq:PropagatorEx1}
\end{align}
Replacing the full propagator in Eq.~(\ref{eq:WilsonFullLines}) with
Eq.~(\ref{eq:PropagatorEx1}), we obtain the following expression
\begin{align}
\mathrm{Tr} W_C & = \left
  \langle\kern-3\nulldelimiterspace\left \langle \mathrm{Tr}\left[ 
S_0^{(1)}(\bm x_1,-T/2, \bm y_1,T/2\; ; a_{I,\bar{I}}) S_0^{(2)}(\bm
  x_2,-T/2, \bm y_2,T/2\; ;a_{I,\bar{I}}) \rangle\right] 
\right\rangle\kern-3\nulldelimiterspace\right \rangle \cr
&-\frac1{4m_Q^2} \int d^4 \eta d^4 \eta' \, \left
  \langle\kern-3\nulldelimiterspace\left \langle \mathrm{Tr}\left[ 
 S_0^{(1)}(\bm x_1,-T/2, \bm \eta,\eta_4\; ; a_{I,\bar{I}})  (-\bm D^2
  +\bm \sigma_1 \cdot \bm B) S_0^{(1)} (\bm \eta,\eta_4, \bm y_1,T/2;a_{I,\bar{I}})
  \right.\right.\right. \cr
&\hspace{1.cm}\times\, \left.\left.\left. S_0^{(2)}(\bm x_2,-T/2, 
  \bm \eta,\eta_4\; ; a_{I,\bar{I}})  (-\bm D^2  +\bm \sigma_2 \cdot \bm B)
  S_0^{(2)} (\bm \eta,\eta_4,\bm y_1,T/2;a_{I,\bar{I}})  
\right]  \right\rangle\kern-3\nulldelimiterspace\right \rangle \cr
&-\frac1{4m_Q^2} 
\left\langle\kern-3\nulldelimiterspace\left \langle \mathrm{Tr}\left[ 
S_0^{(1)}(\bm x_1,-T/2, \bm y_1,T/2\; ;
  a_{I,\bar{I}}) \int d^4 \eta\, S_0^{(2)} (\bm x_2,-T/2,\bm
  \eta,\eta_4;a_{I,\bar{I}})
(\bm   E\cdot \bm D +\bm  \sigma_2 \cdot (\bm E\times \bm D)) 
  \right.\right.\right.\cr 
& \left.\left.\left. \hspace{1.cm} \times \, S_0^{(2)}
  (\bm \eta,\eta_4, \bm y_2,T/2;a_{I,\bar{I}})  \right]
\right\rangle\kern-3\nulldelimiterspace\right \rangle \cr
&-\frac1{4m_Q^2} 
\left\langle\kern-3\nulldelimiterspace\left \langle \mathrm{Tr}\left[ 
\int d^4 \eta\, S_0^{(1)} (\bm x_1-T/2,\bm  \eta,\eta_4;a_{I,\bar{I}}) (\bm
  E\cdot \bm D +\bm  \sigma_1 \cdot (\bm E\times \bm D)) S_0^{(1)}
  (\bm \eta,\eta_4,\bm  y_1,T/2;a_{I,\bar{I}}) 
  \right.\right.\right. \cr   
& \left.\left.\left. \hspace{1.cm} \times \, S_0^{(2)}(\bm x_2,-T/2,
  \bm y_2,T/2\; ;  a_{I,\bar{I}}) \right]
\right\rangle\kern-3\nulldelimiterspace\right \rangle \cr
&+ \frac1{4m_Q^2}  \left\langle\kern-3\nulldelimiterspace\left \langle
  \mathrm{Tr}\left[ S_0^{(1)} (\bm x_1-T/2,\bm y_1,T/2;a_{I,\bar{I}})
\int d^4 \eta d^4 \eta' \,S_0^{(2)}(\bm x_2,-T/2,\bm \eta, \eta_4;a_{I,\bar{I}}) 
 (-\bm D^2 +\bm \sigma_2 \cdot \bm B) \right.\right.\right. \cr
& \left.\left.\left. \hspace{1.cm} \times  S_0^{(2)} (\bm \eta, \eta_4,\bm
  \eta',\eta_4';a_{I,\bar{I}})  (-\bm D^2 +\bm \sigma_2  \cdot \bm B)
  S_0^{(2)} (\bm \eta',\eta_4',\bm y_2,T/2;a_{I,\bar{I}}) \right] 
\right\rangle\kern-3\nulldelimiterspace\right \rangle \cr
&+ \frac1{4m_Q^2}  \left\langle\kern-3\nulldelimiterspace\left \langle
  \mathrm{Tr}\left[ \int d^4 \eta d^4 \eta'\, S_0^{(1)}(\bm x_1,-T/2,\bm \eta,
  \eta_4;a_{I,\bar{I}}) (-\bm D^2 +\bm \sigma_1 \cdot \bm B)
S_0^{(1)}  (\bm \eta,\eta_4,\bm \eta',\eta_4';a_{I,\bar{I}})
  \right.\right.\right. \cr 
&\left.\left.\left.  \hspace{1.cm} \times  \,  (-\bm D^2 +\bm \sigma_1 
  \cdot \bm B) S_0^{(1)}   (\bm \eta', \eta_4',,\bm y_1,T/2;a_{I,\bar{I}})
  S_0^{(2)} (\bm x_2,-T/2,\bm y_2,T/2;a_{I,\bar{I}}) \right] 
\right\rangle\kern-3\nulldelimiterspace\right \rangle.
\label{eq:mQCorrections}
\end{align}
Note that here we consider only the spin-dependent parts. For example,
we can exclude the spin-independent term $\bm D^2/2m_Q$, which is just
the kinetic energy, and that proportional to $\bm \sigma  
\cdot \bm B$, which disappears because of parity
invariance~\cite{Eichten:1980mw}. We can further simplify
Eq.~(\ref{eq:mQCorrections}), leaving all spin-independent parts out,
which are just part of relativistic corrections to the
potential. Taking only the spin-dependent parts into account, we obtain  
\begin{align}
\mathrm{Tr} W_C^{1/m_Q^2} & = 
-\frac1{4m_Q^2} \int d^4 \eta d^4 \eta' \, \left
  \langle\kern-3\nulldelimiterspace\left \langle \mathrm{Tr}\left[ 
 S_0^{(1)}(\bm x_1,-T/2, \bm \eta,\eta_4\; ; a_{I,\bar{I}})  (-\bm
 D^2 ) S_0^{(1)} (\bm \eta,\eta_4, \bm 
  y_1,T/2;a_{I,\bar{I}}) 
  \right.\right.\right. \cr
&\hspace{1.cm}\times\, \left.\left.\left. S_0^{(2)}(\bm x_2,-T/2, 
  \bm \eta,\eta_4\; ; a_{I,\bar{I}})  (\bm \sigma_2 \cdot
  \bm B)   S_0^{(2)} (\bm \eta,\eta_4,\bm y_1,T/2;a_{I,\bar{I}})  
\right]  \right.\right.\cr
&\hspace{2.8cm} + \left.\left.
\mathrm{Tr}\left[  S_0^{(1)}(\bm x_1,-T/2, \bm \eta,\eta_4\; ;
  a_{I,\bar{I}})  (\bm \sigma_1 \cdot \bm B) S_0^{(1)} (\bm
  \eta,\eta_4, \bm  y_1,T/2;a_{I,\bar{I}}) 
  \right.\right.\right. \cr
&\hspace{1.cm}\times\, \left.\left.\left. S_0^{(2)}(\bm x_2,-T/2, 
  \bm \eta,\eta_4\; ; a_{I,\bar{I}})  (-\bm D^2)   S_0^{(2)} (\bm
  \eta,\eta_4,\bm y_1,T/2;a_{I,\bar{I}})   
\right]  \right\rangle\kern-3\nulldelimiterspace\right \rangle \cr
&\hspace{2.8cm} + \left.\left.
\mathrm{Tr}\left[  S_0^{(1)}(\bm x_1,-T/2, \bm \eta,\eta_4\; ;
  a_{I,\bar{I}})  (\bm \sigma_1 \cdot \bm B) S_0^{(1)} (\bm
  \eta,\eta_4, \bm    y_1,T/2;a_{I,\bar{I}}) 
  \right.\right.\right. \cr
&\hspace{1.cm}\times\, \left.\left.\left. S_0^{(2)}(\bm x_2,-T/2, 
  \bm \eta,\eta_4\; ; a_{I,\bar{I}})  (\bm \sigma_2 \cdot
  \bm B)   S_0^{(2)} (\bm \eta,\eta_4,\bm y_1,T/2;a_{I,\bar{I}})  
\right]  \right\rangle\kern-3\nulldelimiterspace\right \rangle \cr
&-\frac1{4m_Q^2} 
\left\langle\kern-3\nulldelimiterspace\left \langle \mathrm{Tr}\left[ 
S_0^{(1)}(\bm x_1,-T/2, \bm y_1,T/2\; ;
  a_{I,\bar{I}}) \right.\right.\right.\cr
&\left.\left.\left. \hspace{1.cm} \times \,\int d^4 \eta\, S_0^{(2)}
  (\bm x_2,-T/2,\bm   \eta,\eta_4;a_{I,\bar{I}})
(\bm  \sigma_2 \cdot (\bm E\times \bm   D)) 
S_0^{(2)}  (\bm \eta,\eta_4, \bm y_2,T/2;a_{I,\bar{I}})  \right]
\right\rangle\kern-3\nulldelimiterspace\right \rangle \cr
&-\frac1{4m_Q^2} 
\left\langle\kern-3\nulldelimiterspace\left \langle \mathrm{Tr}\left[ 
\int d^4 \eta\, S_0^{(1)} (\bm x_1-T/2,\bm  \eta,\eta_4;a_{I,\bar{I}})
  (\bm  \sigma_1 \cdot (\bm E\times \bm D)) S_0^{(1)} 
  (\bm \eta,\eta_4,\bm  y_1,T/2;a_{I,\bar{I}}) 
  \right.\right.\right. \cr   
& \left.\left.\left. \hspace{1.cm} \times \, S_0^{(2)}(\bm x_2,-T/2,
  \bm y_2,T/2\; ;  a_{I,\bar{I}}) \right]
\right\rangle\kern-3\nulldelimiterspace\right \rangle.
\label{eq:WilsonMq2}
\end{align}
The final expression for $W_C^{1/m_Q^2}$ contains $1/m_Q^2$, so that
we can expand the exponential of Eq.~(\ref{eq:WilonPot1}) in powers of
$1/m_Q^2$. Then, Eq.~(\ref{eq:WilsonMq2}) will lead to the
spin-dependent parts of the heavy-quark potential from the instanton
vacuum. The derivation of the potential from Eq.~(\ref{eq:WilsonMq2})
is lengthy but straightforward. In Ref.~\cite{Eichten:1980mw}, it was 
shown in very detail how one can obtain the spin-dependent parts of
the heavy-quark potential in QCD. Since the form of
Eq.~(\ref{eq:WilsonMq2}) is very similar to the corresponding one in
Ref.~\cite{Eichten:1980mw}, we will closely follow the method of
Ref.~\cite{Eichten:1980mw} and refer to it. The leading-order
propagator given in Eq.~(\ref{eq:LO_propagator}) is identified as the
path-order exponential along the time direction apart from the Dirac
delta function. Using the identities for the path-ordered exponentials
given in Appendix, we can proceed to compute each term in 
Eq.~(\ref{eq:WilsonMq2}). Note that the instanton satisfies the
self-duality condition $G_{\mu\nu}^a = \pm \tilde{G}_{\mu\nu}^a$ ($\bm 
B=\pm \bm E$), which plays an essential role in deriving the
spin-dependent potential from the instanton vacuum.  It makes it
possible to relate several independent potentials to the central
potential given in Eq.(\ref{eq:LO_pot}). As a result, all the
spin-dependent potentials are expressed in terms of the central
potential  
\begin{align}
V_{SD}(\bm r) &= \frac1{4m_Q^2}\left(\bm L_1\cdot \bm
  \sigma_2-\bm L_2\cdot \bm \sigma_1 \right) \frac{1}{r}\frac{dV_C(r)}{dr}  + 
\frac{\bm\sigma_1\cdot\bm\sigma_2}{12m_Q^2} \nabla^2 V_C(r) \cr
& \hspace{1cm} +
\frac{1}{3m_Q^2}(3\bm\sigma_1\cdot\bm n\,\bm\sigma_2\cdot \bm 
  n- \bm\sigma_1\cdot\bm\sigma_2) \left(\frac{1}{r}\frac{d}{dr}
-\frac{d^2}{dr^2}\right)V_C(r), 
\label{eq:VSD}
\end{align}
where $\bm L_i$ and $\bm \sigma_i$ represent respectively the orbital
angular momentum and the Pauli spin operator of the corresponding
heavy quark, $\bm n$ designates the unit radial vector. The potential
$V_C(r)$ denotes the central part of the potential that we already have
shown in Eq.~(\ref{eq:LO_pot}). We want to mention that we have used 
$m_Q=m_{\bar{Q}}$. If one considers two heavy quarks with different
masses, we can simply replace $m_Q^2$ with $m_Q m_{\bar{Q}}$ in
Eq.~(\ref{eq:VSD}).   

The spin-dependent potential $V_{SD}$ can be now decomposed into three
different parts, i.e., the spin-spin interaction $V_{SS}(r)$, the
spin-orbit coupling term $V_{LS}(r)$, and the tensor part $V_T(r)$:  
\begin{align}
\label{eq:SPPot}
V_{Q\bar{Q}}(\bm r) = V_C(r)+V_{SS}(r)
(\bm S_Q\!\cdot\!\bm S_{\bar Q})
+V_{LS}(r)(\bm L\cdot\bm S) +V_{T}(r)\left[
3(\bm S_Q\!\cdot\!\bm n)(\bm S_{\bar Q}\!\cdot\! \bm n)-\bm
  S_Q\cdot\bm S_{\bar Q}\right],
\end{align}
where $\bm S_{Q({\bar{Q}})}$ stands for the spin of a heavy quark
(heavy anti-quark) $\bm S_{Q({\bar{Q}})}=\bm\sigma_{1(2)}/2$, $\bm S$
does their total spin $\bm S=\bm S_1+\bm S_2$, and $\bm L$ 
represents the relative orbital angular momentum $\bm L=\bm L_1-\bm
L_2$. Each potential of Eq.~(\ref{eq:SPPot}) is defined respectively
as  
\begin{align}
V_{SS}(r)=\frac{1}{3m_Q^2}\nabla^2 V_C(r),\;\;\;\;
V_{LS}(r)=\frac{1}{2m_Q^2}\frac1{r}\frac{dV_C(r)}{dr},\;\;\;\;
V_{T}(r)=\frac{1}{3m_Q^2}\left(\frac{1}{r}\frac{dV_C(r)}{dr}
-\frac{d^2V_C(r)}{dr^2}\right).
\end{align}
Thus, all three components of the spin-dependent potential are
expressed in terms of the central potential $V_C(r)$. 
\section{Numerical calculations, results and
  discussions \label{sect:Results}} 
\subsection{Instanton potential}
In the instanton liquid model for the QCD vacuum, we have two
important parameters, i.e., the average size of the instanton
$\bar{\rho}\simeq 0.33\,\mathrm{fm}$ and the average distance
$\bar{R}\simeq 1\,\mathrm{fm}$ between instantons, as we have already
mentioned. These numbers were first proposed by
Shuryak~\cite{Shuryak:1981ff} within the instanton liquid model and
were derived from $\Lambda_{\overline{MS}}$ by Diakonov and
Petrov~\cite{Diakonov:1983hh}. Thus, it is also of great interest to
look into the dependence of the heavy-quark potential from the
instanton vacuum on these parameters. Moreover, the values given above 
should not be considered as the exact ones. 
For example, Refs.~\cite{Kim:2005jc, Goeke:2007bj,Goeke:2007nc}
considered $1/N_c$  meson-loop contributions in the light-quark sector
and found it necessesary to readjust the values of parameters as
$\bar{\rho}\simeq0.35\,\mathrm{fm}$ and
$\bar{R}\simeq0.856\,\mathrm{fm}$.  Lattice simulations of the
instanton vacuum suggested $\bar\rho \approx 0.36\,\mathrm{fm}$
and $\bar{R} \approx 0.89\,  \mathrm{fm}$~\cite{Chu:1994vi,
  Negele:1998ev, DeGrand:2001tm,Faccioli:2003qz}, which is almost the
same as those with the $1/N_c$ meson-loop corrections.    
Thus, we want to examine the dependence of the heavy-quark potential
from the instanton vacuum on three different sets of parameters,
that is, Set I~\cite{Diakonov:1983hh,Shuryak:1981ff}, Set
IIa~\cite{Kim:2005jc, Goeke:2007bj, Goeke:2007nc}, and Set
IIb~\cite{Chu:1994vi, Negele:1998ev,
  DeGrand:2001tm,Faccioli:2003qz}.  
The parameter dependence of the potential can be easily understood from 
the form of leading-order potential expressed in Eq.~(\ref{eq:LO_pot}).
While the prefactor $\bar\rho^3/\bar{R}^4N_c$, which includes both 
the parameters, governs the overall strength of the
potential, its range is dictated only by the instanton size $\bar\rho$
through the dimensionless integral $I(r/\bar\rho)$.  

When the quark-antiquark distance is smaller than the instanton size,
i.e., $r\ll \bar{\rho}$ ($x\ll 1$), one can expand the dimensionless
integral $I(x)$ with respect to $x$ 
\begin{align}
I(x)&\simeq \left[\frac{\pi^3}{48}-\frac{\pi^3}{3}J_1(2\pi)\right]x^{2}+\left[
-\frac{\pi^3(438+7\pi^2)}{30720}+\frac{J_2(2\pi)}{80}\right]x^4+{\cal
             O}(x^{6}), 
\end{align}
which yields the central potential in the form of a polynomial 
\begin{equation}
V_C(r)\simeq\frac{4\pi \bar\rho^{\,3}}{\bar{R}^4N_c}\,\left(1.345\,
  \frac{r^2}{\bar\rho^2} -0.501\,\frac{r^4}{\bar\rho^4}\right).
\end{equation}
As the distance between the quark and the antiquark grows larger  
than the intstanton size, i.e. $r\gg \bar{\rho} $ ($x\gg 1$),  
we again get an analytic expression as follows  
\begin{align}
I(x)\simeq-\frac{2\pi^2}{3}\Big[\pi J_0(\pi)+J_1(\pi)\Big] -
  \frac{\pi^2}{2x}+{\cal O}(x^{-2})\,. 
\end{align}
Consequently, the central potential at large $r$ can be approximately 
written as 
\begin{align}
V(r)\simeq 2 \Delta M_Q-\frac{g_{\rm NP}}{r}.
\label{eq:Largex}
\end{align}
The second term behaves like the Coulomb-like potential. So, crudely
speaking, this can be understood as a nonperturbative
contribution to the perturbative one gluon exchange potential from the
instanton vauum at large $r$.  The coupling constant $g_{\mathrm{NP}}$
in Eq.(\ref{eq:Largex}), which is defined as $g_{\rm 
  NP} := 2\pi^3\bar\rho^4/(N_c\bar R^4)$, could be regarded as a
nonperturbative correction to the strong coupling constant
$\alpha_s(r)$. When $r$ goes to infinity  $r\rightarrow\infty$, the
potential is saturated at the value of $2\Delta M_Q$. As discussed
already in Ref.~\cite{Diakonov:1989un}, it implies that the 
instanton vacuum can not explain quark confinement. In the case of
parameter Set~I, which is often considered in the 
light-quark sector, the value of $\Delta M_Q$ is obtained to be
$\Delta M_Q\simeq 66.6\,\mathrm{MeV}$.  However, if one chooses 
Set~IIa, then the result becomes $\Delta M_Q \simeq
143.06\,\mathrm{MeV}$.  The Set~IIb produces $\Delta M_Q \simeq
135.72\,\mathrm{MeV}$. 

\begin{figure*}[htp]
\begin{centering}
\includegraphics[scale=0.75]{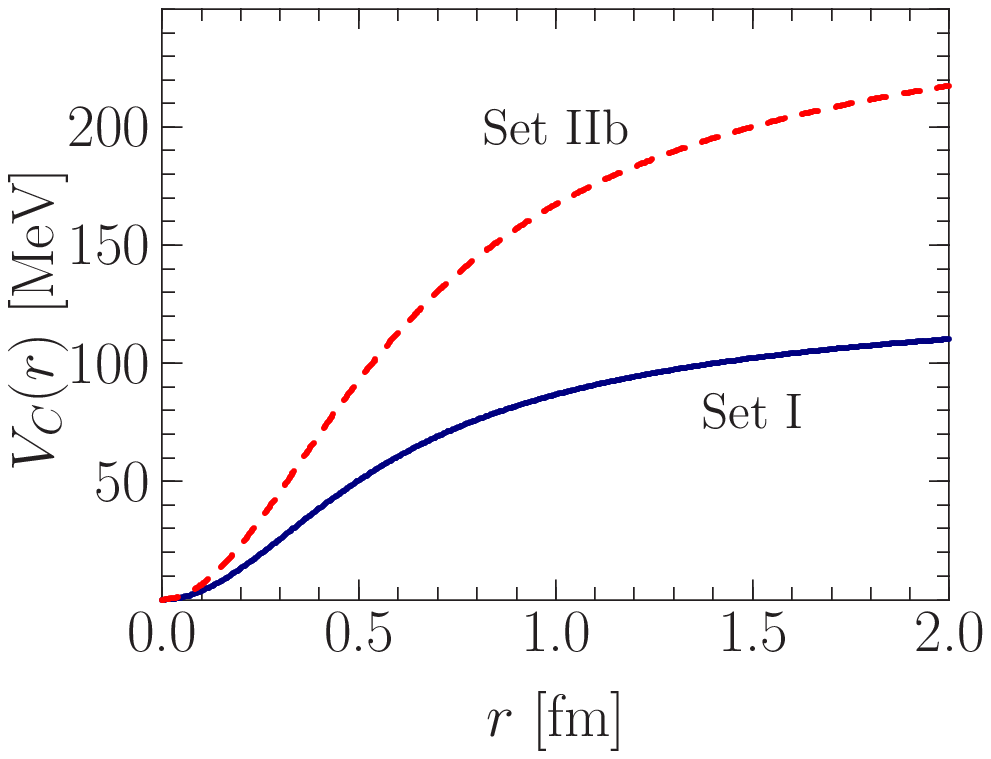}\hspace{0.5cm}
\includegraphics[scale=0.75]{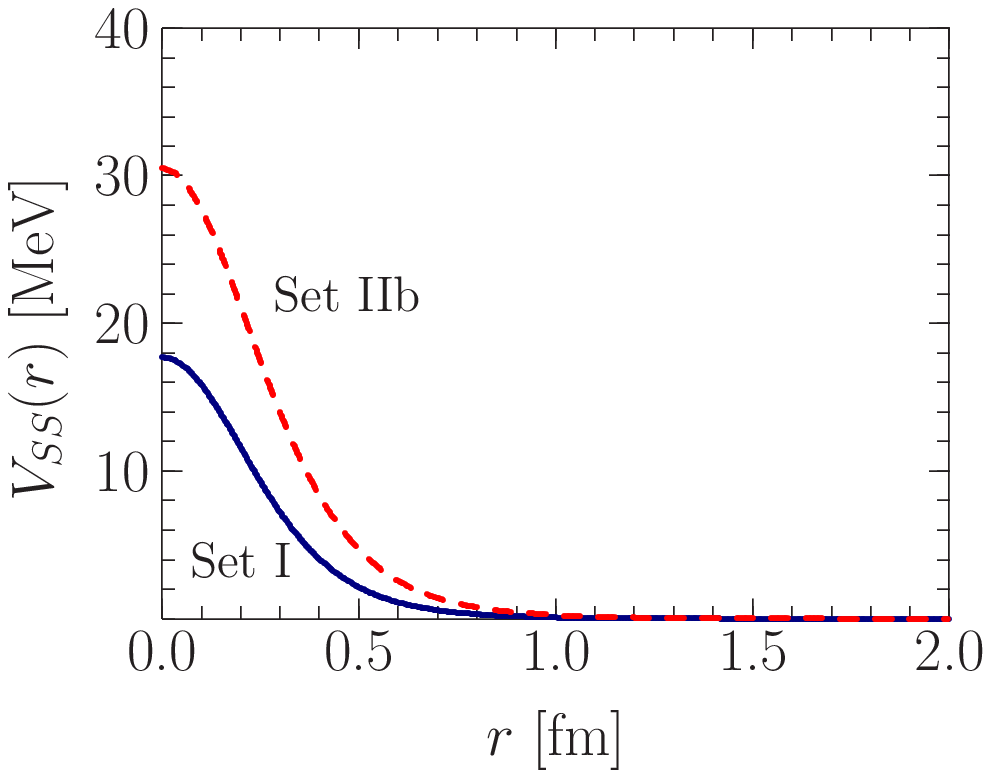}\\
\includegraphics[scale=0.75]{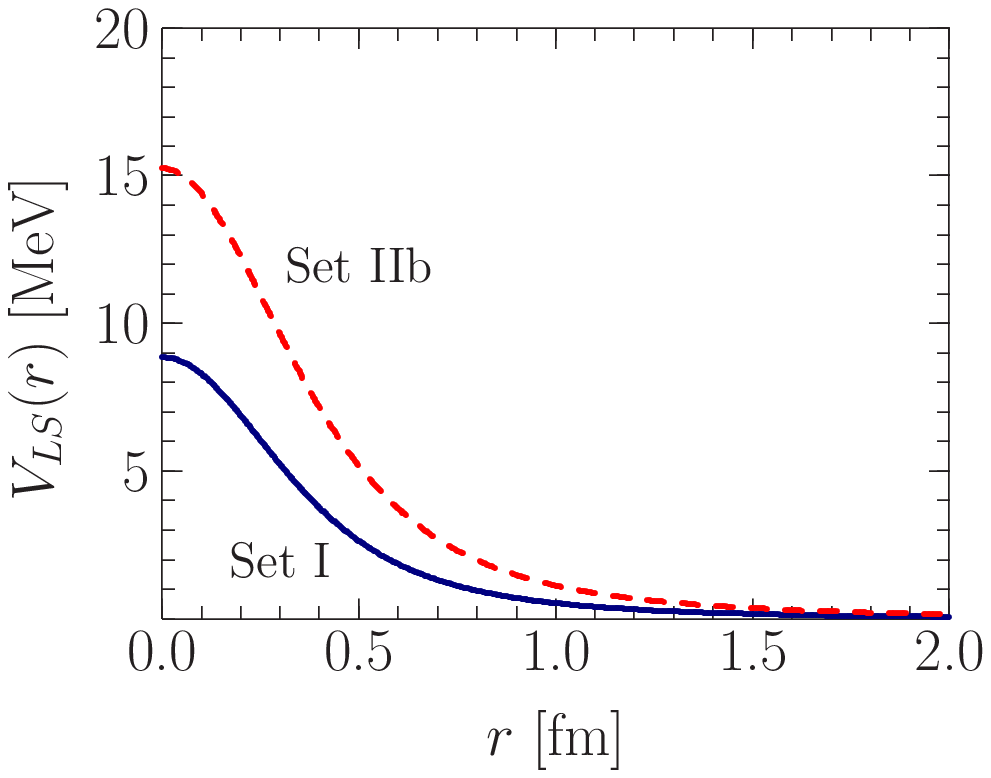}\hspace{0.5cm}
\includegraphics[scale=0.75]{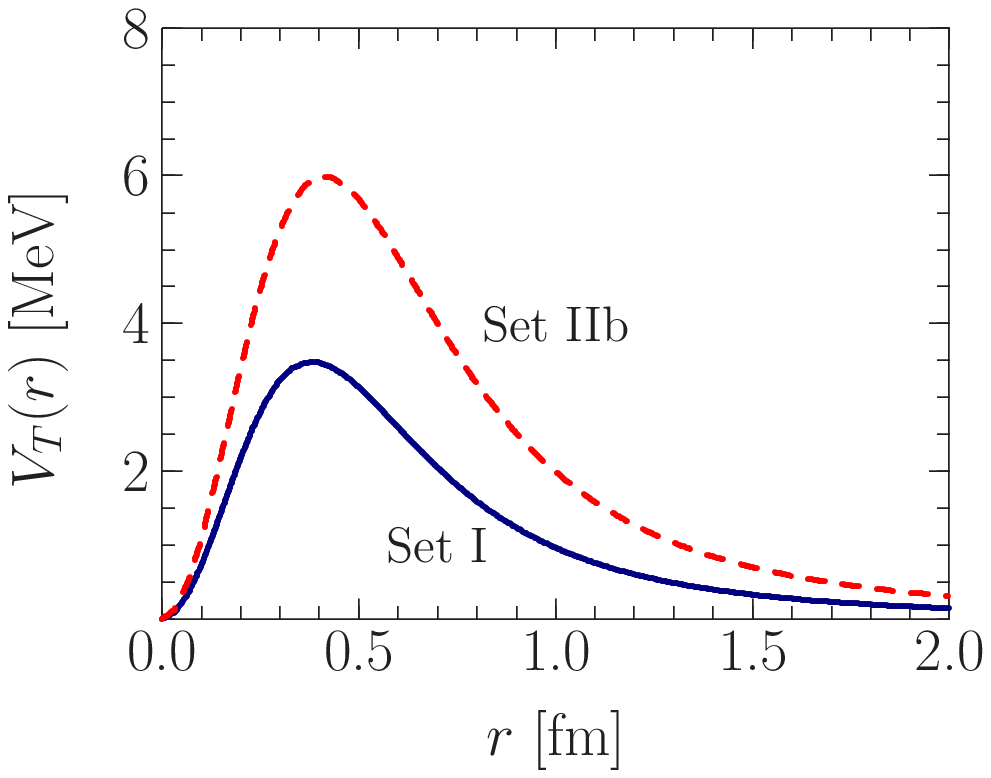}\\
\par\end{centering}
\caption{
Each contribution to the heavy-quark potential as a function
  of $r$ for the two different sets of the instanton parameters
  $\bar{\rho}$ and $\bar{R}$. The upper-left panel depicts the
  central part of the potential, the upper-right panel draws the
  spin-spin interaction, the lower-left panel illustrates the
  spin-orbit part, and the lower-right one shows the tensor
  interaction. The solid curve corresponds to Set~I  with
  $\rho=0.33$~fm and $\bar{R}=1\,\mathrm{fm}$ from the 
 phenomenolgy~\cite{Shuryak:1981ff,Diakonov:1983hh}, whereas   
the dashed one corresponds to Set~IIb with 
  $\rho=0.36$~fm  and $\bar{R}=0.89\,\mathrm{fm}$ 
  from the lattice simulations~\cite{Chu:1994vi, Negele:1998ev,
    DeGrand:2001tm, Faccioli:2003qz}. The mass of the charm quark is
  chosen to be $m_c=1275$~MeV.} 
\label{fig2}
\end{figure*}
Figure~\ref{fig2} draws $r$ dependence of each term of the
heavy-quark potentials from the instanton vacuum.  We take into
account the charm quark sector as an example. We also show the
dependence of each term of the potential on two different 
sets of parameters, that is, Set~I and Set~IIb. 
One can see that the central part of the potential increases
monotonically at small distances $r\ll\bar{\rho}$ and later becomes almost 
linear at the distances comparable with the instanton size $r\sim\bar\rho$
as already discussed in Ref.~\cite{Diakonov:1989un}.
At large $r\gg \bar{\rho}$ it starts  to get saturated at the value 
$V_C(r\rightarrow\infty)\simeq 133.2$\,MeV with Set~I
and $V_C(r\rightarrow\infty)\simeq 271.44$\,MeV with Set~IIb.
The spin-spin interaction part is of particular interest among
these contributions to the spin-dependent potential. In pQCD, it is
given as a point-like interaction~\cite{Brambilla:2004jw} in the
leading order. On the other hand, the spin-spin interaction from the
instanton vacuum looks similar to a Gaussian-type interaction. The
spin-orbit potential behaves in a similar way to the spin-spin
potential. The tensor interaction, however, shows a different $r$
dependence. As $r$ increases, the tensor potential vanishes at $r=0$
and then starts to increase until $r\approx 0.4\,\mathrm{fm}$, from
which it begins to fall off. The strength of each part of the
potential become stronger when smaller value of $\bar{R}$ is employed,
since all terms turn out to be very sensitive to $\bar{R}$ on account
of the prefactor $\bar\rho^3/\bar{R}^4N_c$. It implies that a less
dilute instanton medium yields stronger interactions between a heavy
quark and a heavy antiquark. However, one has to keep in mind that the
value of $\bar{R}$ should not be continually decreased, because the
whole framework of the instanton liquid model is based on the
diluteness of the instanton medium where the packing parameter
proportional to $(\bar{\rho}/\bar{R})^4$ must be kept as a small
parameter.     

On the other hand,
the change of the $\bar{\rho}$ value seems less effective in the
spin-dependent parts of the potential. 
This is again due to the fact that all spin dependent parts have 
the prefactor 
$\bar\rho/\bar{R}^4N_c$ where instanton size appears in the first order,
after rewriting the spin-dependent parts in terms of the dimensionless integral $I(x)$ 
(see Eq.~(\ref{dimint})) and its derivatives.
As mentioned in the previous section, the
average size of the instanton has a physical meaning of the
renormalization scale~\cite{Diakonov:1985eg, Son:2015bwa}, which 
is a crucial virtue of the instanton liquid model.  Thus,
$\bar{\rho}=0.33\,\mathrm{fm}$ indicates the renormalization scale 
$\mu=600\,\mathrm{MeV}$. Bearing in mind this meaning of $\bar{\rho}$, 
we should not take the value of $\bar{\rho}$ freely.  
Note that the value of $\bar{\rho}^{-1} = 600\,\mathrm{MeV}$ implies
the strong couipling constant frozen at $\bar{\rho}^{-1}$. 
Thus, Fig.~\ref{fig2}  shows the dependence of the
heavy-quark potential on both $\bar{\rho}$ and $\bar{R}$ within a
constraint range of their values. 
In the case of the bottom quarks and anti-quarks, the instanton
effects are quite much suppressed because of the large bottom quark mass. 

For completeness, we provide the expression for the matrix elements
of  $Q\bar{Q}$ potential in Eq.~(\ref{eq:SPPot}) 
\begin{align}
\langle {}^{2S+1}L_J| V_{Q\bar{Q}}(\bm{r}) | {}^{2S+1}L_J\rangle &= V(r)+
\left[ \frac12S(S+1)-\frac34\right] V_{SS}(r)  +  \frac12\left[J(J+1) 
  -L(L+1) -S(S+1)\right]    V_{LS}(r)  \cr
&\left\{-\frac{3\left[J(J+1) 
  -L(L+1) -S(S+1)\right]
  \left[J(J+1) 
  -L(L+1) -S(S+1)+1\right]}{2(2L-1)(2L+3)}\right.\cr
&\left.+\frac{2S(S+1)L(L+1)}{(2L-1)(2L+3)}\right\}
V_T(r),
\label{eq:PotMatrix}
\end{align}
where we have used the conventional spectroscopic notation
${}^{2S+1}L_J$ given in terms of  the total spin $S$, the orbital
angular momentum $L$, and the total angular momentum $J$ satisfying
the relation $\bm J= \bm L+\bm S$. 
\subsection{Gaussian Expansion Method}
In order to evaluate the bound states in the spectrum of quarkonia, we
need to solve the Schr\"{o}dinger equation with the potential from the
instanton vacuum given in Eq.~(\ref{eq:SPPot})
\begin{align}
\left[-\frac{\hbar^2}{m_Q}\nabla^2+V_{Q\bar
  Q}(\bm{r})-E\right]\Psi_{JM}(\bm{r})=0, 
\label{eq:Schroe}
\end{align}
where $m_Q$ arises from the doubled reduced mass of the quarkonium
system and $\Psi_{JM}$ represents the wave function of the state with
the total angular momentum $J$ and its third component $M$. We can
solve Eq.(\ref{eq:Schroe}) numerically, using the  
Gaussian expansion method~(see review~\cite{Hiyama2003}) in which
the wave function is expanded in terms of a set of $L^2$-integrable basis 
functions $\{\Phi_{JM,n}^{LS};\,n=1-n_{\mathrm{max}}\}$
\begin{align}
\Psi_{JM}(\bm{r})=\sum_{n=1}^{n_{\rm max}}C_{n,LS}^{(J)}\Phi_{JM,n}^{LS}(\bm{r})
\end{align}
and the Rayleigh-Ritz variational principle is employed. Thus, one can
formulate a generalized eigenvalue problem given as 
\begin{align}
\sum_{m=1}^{n_{\rm max}}\langle\Phi_{JM,n}^{LS}\left|-\frac{\hbar^2}{m_Q}\nabla^2
+V_{Q\bar Q}(\bm r)-E\right|\Phi_{JM,m}^{LS}\rangle C_{m,LS}^{(J)}=0\,.
\end{align}
The normalized radial part of the basis wave functions
$\phi_{n}^{L}(r)$ is expressed in terms of the Gaussian basis functions 
\begin{align}
\phi_{n}^{L}(r)=\left(\frac{2^{2L+\frac{7}{2}}r_{n}^{-2L-3}}{\sqrt{\pi}(2L+1)!!}\right)^{1/2}
r^{L}e^{-(r/r_{n})^2},
\end{align}
where $r_n,\, n=1,2,...,n_{max}$ stand for variational parameters. 
In the case of a two-body problem, the total number of the variational
parameters can be reduced by choosing the geometric progression in the
form of $r_n=r_1a^{n-1}$, which produces a good convergence of the
results. Thus, we need only three variational parameters, i.e. $r_1$, $a$
and $n_{\mathrm{max}}$.

\subsection{Quarkonium states}
We already mentioned that at large distance the instanton potential is
saturated, so that there is no confinement in the present approach. The
bound or quasibound charmonium states with the masses below or around
the threshold mass $M_{Q\bar{Q}}\simeq 2(m_c+\Delta M_Q)$,
where $m_c=1275$ is the charm quark mass~\cite{PDG2016},  
are listed in Table~\ref{table1} with the two different sets of the
instanton parameters. Other states above threshold will appear
as resonances in the present approach. 
\begin{table}[hbt]
\begin{ruledtabular}
\begin{tabular}{l|cc|c}
&This work &{This work}& \\
&Set~I& Set~IIb & Experiment~\cite{PDG2016}\\
&$\bar\rho=1/3$\,fm, $R=1$\,fm~\cite{Shuryak:1981ff,Diakonov:1983hh}
&\qquad\qquad$\bar\rho=0.36$\,fm, 
$R=0.89$\,fm~\cite{Chu:1994vi, Negele:1998ev,
  DeGrand:2001tm,Faccioli:2003qz}\qquad\qquad&[MeV]\\ 
&[MeV]& [MeV]\\
\hline
$M_{\eta_c}$ & 2668.81 & 2753.64 & $2983.6\pm 0.6$\\
$M_{J/\psi}$ & 2669.57 & 2755.36 & $3096.916\pm 0.11$\\ 
$M_{\chi_{c0}}$ &  2692.43 &  2800.86 & $3414.75\pm0.31$\\
$M_{\chi_{c1}}$ & 2692.50 &  2801.11 & $3510.66\pm0.07$\\
$M_{\chi_{c2}}$ &  2692.67 &  2801.70 &$3556.20\pm 0.09$\\
\end{tabular}
\end{ruledtabular}
\caption{Low-lying charmonium states from the instanton potential. 
Charm quark mass is set to be $m_c=1275$~MeV.
}
\label{table1}
\end{table}
One can see that the instanton effects are not small in reproducing the 
mass of quarkonia. For example, in the case of the potential with parameter Set~I, 
the contribution to the mass of a charmonium is determined by $\Delta
M_{c\bar{c}}=M_{c\bar{c}}-2m_c$. For example, the contribution of the
instanton effects to the $\eta_c$ mass turns out to be
$118.81\,\mathrm{MeV}$, which is approximately about $30\,\%$ in
comparison with the experimental data~433.60\,MeV. As discussed
already, the potential from the instanton vacuum is sensitive to the
instanton parameters.  Therefore, the change in the instanton
parameters strongly affects the spectrum of $Q\bar{Q}$ states. For
example, parameter Set~IIb gives the result $\Delta M_{\eta_c}\simeq
203.64$\,MeV, which is almost $50\,\%$, compared to the
data. Parameter Set~IIa gives slightly larger results than those with
Set IIb. When it comes to the $J/\psi$ state, the instanton effects on the
$Q\bar{Q}$  mass becomes smaller in comparison with the experimental
data. However, it is still important to consider them, since 
$\Delta M_{J/\psi}$ is 119.57\,MeV (205.36\,MeV) with Set I (Set IIb)
used, compared with the data 540.92\,MeV. On the other hand, 
we obtain $\Delta M_{\chi_{c0}}\simeq 142.43$\,MeV (Set I) and  $\Delta
M_{\chi_{c0}}\simeq 250.86$\,MeV (Set~IIb).
Parameter Set~I reproduces $\chi_{c0}$, $\chi_{c1}$ and $\chi_{c2}$ as
quasibound  states while parameter Set~IIb yields them as the
definite bound states.   

It is of also interest to discuss the effects of the hyperfine mass
splitting from the instanton vacuum. The contribution to the hyperfine
mass splitting of each low-lying charmonium state is listed in 
Table~\ref{table2}.  
\begin{table}[hbt]
\begin{ruledtabular}
\begin{tabular}{l|cc|c}
&This work &{This work}& \\
&Set~I& Set~IIb & Experiment ~\cite{PDG2016}\\
&$\bar\rho=1/3$\,fm, $R=1$\,fm~\cite{Shuryak:1981ff,Diakonov:1983hh}
&\qquad\qquad$\bar\rho=0.36$\,fm, 
$R=0.89$\,fm~\cite{Chu:1994vi, Negele:1998ev,
  DeGrand:2001tm,Faccioli:2003qz}\qquad\qquad&[MeV]\\ 
&[MeV]& [MeV]\\
\hline
$\Delta M_{J/\psi-\eta_c}$          & 0.72  & 1.72 & 113.32 $\pm$ 0.70\\ 
$\Delta M_{\chi_{c1}-\chi_{c0}}$ &  0.07 & 0.25 & \,\,\,95.91 $\pm$ 0.32\\
$\Delta M_{\chi_{c2}-\chi_{c0}}$ &  0.24 & 0.84 & 141.45 $\pm$ 0.32\\
$\Delta M_{\chi_{c2}-\chi_{c1}}$ &  0.16 & 0.59 &  \,\,\,45.54 $\pm$ 0.11\\
\end{tabular}
\end{ruledtabular}
\caption{Contributions to the hyperfine mass splittings of the low-lying charmonium states. 
Charm quark mass is set to be $m_c=1275$~MeV.}
\label{table2}
\end{table}
While the instanton effects come into play significantly on $\Delta
M_{c\bar{c}}$, they turn out to be rather small in describing the hyperfine mass
splittings of the charmonia. This might be due to the fact that the
spin-dependent part of the potential from the instanton vacuum is
almost an order of magnitude smaller than the central part. 
The tensor interaction almost does not contribute to the results.
As a result,  the instanton effects on the hyperfine mass splittings 
are almost negligible. In order to obtain realistic results of the
hyperfine mass splittings as well as of the charmonium masses, 
we need to include the Coulomb-like potential coming from the
perturbative one gluon-exchange  and the confining potential together
with that from the instanton vacuum. 

\begin{table}[hbt]
\begin{ruledtabular}
\begin{tabular}{l|cc}
&This work & \\
&Set~I&  Experiment~\cite{PDG2016}\\
&$\bar\rho=1/3$\,fm, $R=1$\,fm~\cite{Shuryak:1981ff,Diakonov:1983hh}
 &[MeV]\\ &[MeV]& \\
\hline
$M_{\eta_b}$ & 8454.58 &  $9399.0\pm 2.3$\\
$M_{\Upsilon}$ & 8454.76 &  $9460.30\pm 0.26$\\ 
$M_{\chi_{b0}}$ &  8477.95 & $9859.44\pm0.52$\\
$M_{\chi_{b1}}$ & 8477.97  & $9892.78\pm0.40$\\
$M_{\chi_{b2}}$ &  8478.01 &$9912.21\pm 0.40$\\
\end{tabular}
\end{ruledtabular}
\caption{Low-lying bottomonium states from the instanton potential. 
Charm quark mass is set to be $m_b=4180$~MeV.
}
\label{table3}
\end{table}

\section{Summary and outlook \label{sect:Summary}}
In the present work, we aimed at investigating the instanton effects
on the heavy-quark potential, based on the instanton liquid model. We
first considered the heavy-quark propagator starting from the QCD
Lagrangian, which comes into an essential play in deriving the
heavy-quark potential. We showed briefly how to construct 
the heavy-quark potential from the instanton vacuum. Expanding the
heavy-quark propagator in powers of the inverse mass of the heavy
quark, we obtained the spin-dependent parts of the heavy-quark
potential. We studied the dependence of the heavy-quark potential on
the two essential parameters for the instanton vacuum, that is, the
average size of the instanton $(\bar{\rho})$ and the inter-distance
between the instantons $(\bar{R})$. The results of the potential are
very sensitive to the parameter $\bar{R}$, while they are varied 
marginally with $\bar{\rho}$ changed. The spin-spin interaction shows
$r$ dependence similar to a Gaussian-type potential, which is
distinguished from the point-like spin-spin interaction derived from 
perturbative QCD. The spin-orbit potential behaves like the
spin-spin interaction, whereas the tensor potential exhibits a different
character. It increases until $r$ reaches approximately $0.4$ fm and
then starts to fall off.  

Having solved explicitly the Schr\"{o}dinger equation with the
heavy-quark potential purely induced by the instantons, we discussed   
the masses of the low-lying quakonia. The instanton contribution to
the hyperfine mass splitting turns out to be tiny due to smallness of
the spin-dependent part of the potential. We also discussed the
dependence of the results on the intrinsic parameters of the instanton
vacuum, i.e. the average size of the instanton and the inter-distance
between instantons.     

It is of great importance to study carefully the mass spectra
of the quarkonia and their decays by solving explicitly the
Schr\"odinger equation, combining the heavy-quark potential derived 
in the present work with the confining and Coulomb
potentials. Considering the fact that the instanton vacuum plays a 
key role in realizing chiral symmetry and its spontaneous breaking in
QCD, the nonperturbative gluon dynamics is expected to shed light on
strong decays of the quarkonia involving pions. 
Since the central part of the heavy-quark potential was derived by
using the small packing parameter $N/VN_c$, we can obtain the
corrections from the next-to-leading order $(N/VN_c)^2$. In principle,
it is not that difficult to compute them. Starting from the instanton
operator corresponding to the Wilson line (see Eq.(17) in 
Ref.~\cite{Diakonov:1989un}), we can consider the next-to-leading
order in the expansion with respect to the small packing parameter of
the instanton medium. Though the corrections from the next-to-leading
order might be very small, one could use it for the fine-tuning of the
mass spectrum of the quarkonia. The corresponding investigations are 
under way.   
\begin{appendix}
\section{Useful formulae}  
Using the instanton and anti-instanton fields 
\begin{align}
A_{I\mu}=\frac{x_\nu\bar{\eta}_{\mu\nu}^a \tau^a
  \rho^2}{x^2(x^2+\rho^2)},    \qquad
  A_{{\bar I}\mu}=\frac{x_\nu{\eta}_{\mu\nu}^a \tau^a
  \rho^2}{x^2(x^2+\rho^2)}, 
\label{app:1}
\end{align}
where $\bar{\eta}_{\mu\nu}^a$ and ${\eta}_{\mu\nu}^a$  denote the 't Hooft
symbols~\cite{'tHooft:1976fv}, 
we can easily derive the path-ordered exponential as
follows~\cite{Diakonov:1989un}  
\begin{align}
P \exp\left(i\int_{-\infty}^\infty dx_4 A_{I4} \right) =
-\cos \left( \frac{\pi |\bm z|}{\sqrt{\rho^2+z^2}}\right) -
  i\frac{\bm\tau\cdot \bm z}{|\bm z|}\sin\left( \frac{\pi |\bm
  z|}{\sqrt{\rho^2+z^2}}\right), 
\label{app:2}  
\end{align}
which was used for deriving the heavy-quark potential and the
instanton corrections to the heavy-quark mass. 

The leading-order propagator given in Eq.~(\ref{eq:LO_propagator}) is
the same as the path-ordered exponential apart from the Dirac delta
function. Thus, it is of great use to consider the
identities derived in Ref.~\cite{Eichten:1980mw} for the path-order
exponentials when we compute the spin-dependent parts of the
heavy-quark potential. Defining the path-ordered exponential as  
\begin{align}
P(x_4,y_4) := P\exp\left( i\int_{x_4}^{y_4} dz_4 A_4(z) \right),
\end{align}
we have the following identities  
\begin{align}
\label{app:3}
&P(x_4,y_4)P(y_4,z_4) =P(x_4,z_4),\cr
& D_i(x_4) P(x_4,y_4) - P(x_4,y_4) D^i(y_4) = \int_{y_4}^{x_4} dz
  P(x_4,z) E_i(z)P(z,y_4),\cr
& P(\bm y,t;\bm x,t) D_i(\bm x, t) P(\bm x,t;\bm y,t) = D_i(\bm y,t) -
  \epsilon_{ijk} \int_0^1 d\alpha (x-y)_j [P(\bm y,t;\bm z,t) B_k(\bm
  z, t) P(\bm z,t;\bm y,t)],   
\end{align}
where $\bm z=\alpha \bm y +(1-\alpha) \bm x$. $D_i$ denotes the
spatial component of the covariant derivative. When time $t$ goes to
infinity, i.e. $t=\pm\frac{T}{2}\to \infty$, the third identity is
simplified to be 
\begin{align}
  \label{app:4}
\lim_{|t|\to\infty} P(\bm y,t;\bm x,t) \bm D(\bm x, t) P(\bm x,t;\bm
  y,t) = i \nabla_y. 
\end{align}

\end{appendix}

\section*{Acknowledgments}
HChK wants to express his gratitude to A. Hosaka, M. Oka,
and Q. Zhao for very useful comments and discussions at ``The 31st Reimei
Workshop on Hadron Physics in Extreme Conditions at J-PARC''.
HChK owes also debt of thanks to the late D. Diakonov and V. Petrov
for invaluable discussions and suggestions. This work is supported by
the Basic Science Research Program through the National Research
Foundation (NRF) of Korea funded by the Korean government (Ministry of
Education, Science and Technology, MEST), Grant
Numbers~2016R1D1A1B03935053 
(UY) and 2015R1D1A1A01060707~(HChK). The work was also partly
supported by RIKEN iTHES Project.


\begin{thebibliography}{99}
\bibitem{Choi:2003ue} 
  S.~K.~Choi {\it et al.} [Belle Collaboration],
  Phys.\ Rev.\ Lett.,\ {\bf 91}:   262001 (2003)

\bibitem{Aubert:2004ns} 
  B.~Aubert {\it et al.} [BaBar Collaboration],
  Phys.\ Rev.\ D, \ {\bf 71}:  071103 (2005)

\bibitem{Aubert:2005rm} 
  B.~Aubert {\it et al.} [BaBar Collaboration],
  Phys.\ Rev.\ Lett.,\ {\bf 95}:  142001 (2005)

\bibitem{Abe:2007jna} 
  K.~Abe {\it et al.} [Belle Collaboration],
  Phys.\ Rev.\ Lett.,\ {\bf 98}:  082001  (2007)

\bibitem{Choi:2007wga} 
  S.~K.~Choi {\it et al.}  [Belle Collaboration],
  Phys.\ Rev.\ Lett.,\ {\bf 100}:  142001 (2008)

\bibitem{Belle:2011aa} 
  A.~Bondar {\it et al.} [Belle Collaboration],
  Phys.\ Rev.\ Lett.,\ {\bf 108}: 122001 (2012)

\bibitem{Liu:2013dau} 
  Z.~Q.~Liu {\it et al.} [Belle Collaboration],
  Phys.\ Rev.\ Lett.,\ {\bf 110}: 252002 (2013)

\bibitem{Ablikim:2013mio} 
  M.~Ablikim {\it et al.} [BESIII Collaboration], Phys.\ Rev.\ Lett.,\
  {\bf 110}: 252001 (2013) 

\bibitem{Ablikim:2013wzq} 
  M.~Ablikim {\it et al.} [BESIII Collaboration],
  Phys.\ Rev.\ Lett.,\ {\bf 111}:  242001 (2013)
  
\bibitem{Aaij:2013zoa} 
  R.~Aaij {\it et al.} [LHCb Collaboration],
  Phys.\ Rev.\ Lett.,\ {\bf 110}:  222001 (2013)
  
\bibitem{Ablikim:2013xfr} 
  M.~Ablikim {\it et al.} [BESIII Collaboration],
  Phys.\ Rev.\ Lett.,\  {\bf 112}:  022001 (2014)

\bibitem{Aaij:2014jqa}
  R.~Aaij {\it et al.}  [LHCb Collaboration],
  Phys.\ Rev.\ Lett.,\ {\bf 112}:  222002 (2014)

\bibitem{Aaij:2015zxa} 
  R.~Aaij {\it et al.} [LHCb Collaboration],
  Phys.\ Rev.\ D, {\bf 92}:  112009 (2015)

\bibitem{Aubert:2008ba} 
  B.~Aubert {\it et al.} [BaBar Collaboration],
  Phys.\ Rev.\ Lett.,\  {\bf 101}:  071801 (2008)
  [Phys.\ Rev.\ Lett.,\  {\bf 102}:  029901 (2009)]


\bibitem{Aubert:2009as} 
  B.~Aubert {\it et al.} [BaBar Collaboration],
  Phys.\ Rev.\ Lett.,\  {\bf 103}: 161801 (2009)

\bibitem{Bonvicini:2009hs} 
  G.~Bonvicini {\it et al.} [CLEO Collaboration],
  Phys.\ Rev.\ D, {\bf 81}:  031104 (2010)

\bibitem{Mizuk:2012pb} 
  R.~Mizuk {\it et al.} [Belle Collaboration],
  Phys.\ Rev.\ Lett.,\  {\bf 109}: 232002 (2012)

\bibitem{Dobbs:2012zn} 
  S.~Dobbs, Z.~Metreveli, K.~K.~Seth, A.~Tomaradze and T.~Xiao,
  Phys.\ Rev.\ Lett.,\  {\bf 109}:  082001 (2012)

\bibitem{Tamponi:2015xzb} 
  U.~Tamponi {\it et al.} [Belle Collaboration],
  Phys.\ Rev.\ Lett.,\  {\bf 115}:  142001 (2015)

\bibitem{Abdesselam:2015zza} 
  R.~Mizuk {\it et al.} [Belle Collaboration],
  Phys.\ Rev.\ Lett.,\  {\bf 117}: 142001 (2016)

\bibitem{Aaij:2015tga} 
  R.~Aaij {\it et al.} [LHCb Collaboration],
  Phys.\ Rev.\ Lett.,\  {\bf 115}: 072001 (2015)

\bibitem{Bevan:2014iga} 
  A.~J.~Bevan {\it et al.} [BaBar and Belle Collaborations],
  Eur.\ Phys.\ J.\ C, {\bf 74}:  3026 (2014)

\bibitem{Andronic:2015wma} 
  A.~Andronic {\it et al.},
  Eur.\ Phys.\ J.\ C, {\bf 76}: 107 (2016)

\bibitem{Yuan:2015kya} 
  C.~Z.~Yuan [BESIII Collaboration],
  Front.\ Phys.\ China, {\bf 10}: 101401 (2015)

\bibitem{Yuan:2015ztu} 
  C.~Z.~Yuan [Belle Collaboration],
  arXiv:1512.03281 [hep-ex]

\bibitem{Swanson:2006st} 
  E.~S.~Swanson,
  Phys.\ Rept.,\  {\bf 429}:  243 (2006).

\bibitem{Eichten:2007qx} 
  E.~Eichten, S.~Godfrey, H.~Mahlke and J.~L.~Rosner,
  Rev.\ Mod.\ Phys.,\  {\bf 80}: 1161---1193 (2008)

\bibitem{Voloshin:2007dx} 
  M.~B.~Voloshin,
  Prog.\ Part.\ Nucl.\ Phys.,\  {\bf 61}:  455---511 (2008)

\bibitem{Brambilla:2010cs} 
  N.~Brambilla {\it et al.},
  Eur.\ Phys.\ J.\ C, {\bf 71}:  1534 (2011)

\bibitem{Olsen:2014qna} 
  S.~L.~Olsen,
  Front.\ Phys.,\  {\bf 10}: 101401 (2015)

\bibitem{Penin:2009wf} 
  A.~A.~Penin,
  arXiv:0905.4296 [hep-ph].

\bibitem{Recksiegel:2003fm} 
  S.~Recksiegel and Y.~Sumino,
  Phys.\ Lett.\ B, {\bf 578}:  369---375 (2004)

\bibitem{Kniehl:2003ap} 
  B.~A.~Kniehl, A.~A.~Penin, A.~Pineda, V.~A.~Smirnov and
  M.~Steinhauser, 
  Phys.\ Rev.\ Lett.,\  {\bf 92}:  242001 (2004)
  [Phys.\ Rev.\ Lett.,\  {\bf 104}:  242001 (2010)].

\bibitem{Patrignani:2012an} 
  C.~Patrignani, T.~K.~Pedlar and J.~L.~Rosner,
  Ann.\ Rev.\ Nucl.\ Part.\ Sci.,\  {\bf 63}: 21---44 (2013)

\bibitem{Eichten:1974af} 
  E.~Eichten, K.~Gottfried, T.~Kinoshita, J.~B.~Kogut, K.~D.~Lane and
  T.~M.~Yan, 
  Phys.\ Rev.\ Lett.,\  {\bf 34}:  369---372 (1975)

\bibitem{Eichten:1978tg} 
  E.~Eichten, K.~Gottfried, T.~Kinoshita, K.~D.~Lane and T.~M.~Yan,
  Phys.\ Rev.\ D, {\bf 17}:  3090---3117 (1978)
  [Phys.\ Rev.\ D, {\bf 21}: (1980) 313].

\bibitem{Susskind:1976pi} 
  L.~Susskind,
``Coarse Grained Quantum Chromodynamics,'' in 
Weak and Electromagnetic Interactions at high energies:
Proceedings. Edited by Roger Balian and Christopher H. Llewellyn 
Smith (N.Y., North-Holland, 1977).

\bibitem{Appelquist:1977tw} 
  T.~Appelquist, M.~Dine and I.~J.~Muzinich,
  Phys.\ Lett.\ B, {\bf 69}:  231---236 (1977)

\bibitem{Appelquist:1977es} 
  T.~Appelquist, M.~Dine and I.~J.~Muzinich,
  Phys.\ Rev.\ D, {\bf 17}:  2074---2081 (1978)

\bibitem{Fischler:1977yf} 
  W.~Fischler,
  Nucl.\ Phys.\ B, {\bf 129}:  157---174 (1977)

\bibitem{Peter:1996ig} 
  M.~Peter,
  Phys.\ Rev.\ Lett.,\  {\bf 78}: 602---605 (1997)

\bibitem{Peter:1997me} 
  M.~Peter,
  Nucl.\ Phys.\ B, {\bf 501}:  471---494 (1997)

\bibitem{Schroder:1998vy} 
  Y.~Schr\"oder,
  Phys.\ Lett.\ B, {\bf 447}:  321---326 (1999)

\bibitem{Smirnov:2009fh} 
  A.~V.~Smirnov, V.~A.~Smirnov and M.~Steinhauser,
  Phys.\ Rev.\ Lett.,\  {\bf 104}: 112002 (2010)

\bibitem{Anzai:2009tm} 
  C.~Anzai, Y.~Kiyo and Y.~Sumino,
  Phys.\ Rev.\ Lett.,\  {\bf 104}: 112003 (2010)

\bibitem{Wilson:1974sk} 
  K.~G.~Wilson,
  Phys.\ Rev.\ D, {\bf 10}:  2445---2459 (1974)

\bibitem{Bali:1992ab} 
  G.~S.~Bali and K.~Schilling,
  Phys.\ Rev.\ D, {\bf 46}:  2636---2646 (1992)

\bibitem{Booth:1992bm} 
  S.~P.~Booth {\it et al.} [UKQCD Collaboration],
  Phys.\ Lett.\ B, {\bf 294}: 385---390 (1992)

\bibitem{Bali:1996cj} 
  G.~S.~Bali, K.~Schilling and A.~Wachter,
  Phys.\ Rev.\ D, {\bf 55}:  5309---5324 (1997)

\bibitem{Glassner:1996xi} 
  U.~Glassner {\it et al.} [SESAM Collaboration],
  Phys.\ Lett.\ B, {\bf 383}: 98---104 (1996)

\bibitem{Bali:1997am} 
  G.~S.~Bali, K.~Schilling and A.~Wachter,
  Phys.\ Rev.\ D, {\bf 56}: 2566---2589 (1997)

\bibitem{Bali:2000gf} 
  G.~S.~Bali,
  Phys.\ Rept.,\  {\bf 343}: 1---136 (2001)

\bibitem{Kawanai:2013aca} 
  T.~Kawanai and S.~Sasaki,
  Phys.\ Rev.\ D, {\bf 89}: 054507 (2014) 

\bibitem{Kawanai:2015tga} 
  T.~Kawanai and S.~Sasaki,
  Phys.\ Rev.\ D, {\bf 92}:  094503 (2015)

\bibitem{Belavin:1975fg} 
  A.~A.~Belavin, A.~M.~Polyakov, A.~S.~Schwartz and Y.~S.~Tyupkin,
  Phys.\ Lett.\ B, {\bf 59}: 85---87 (1975)

\bibitem{Wilczek:1977md} 
  F.~Wilczek and A.~Zee,
  Phys.\ Rev.\ Lett.,\  {\bf 40}: 83---86 (1978)

\bibitem{Callan:1978ye} 
  C.~G.~Callan, Jr., R.~F.~Dashen, D.~J.~Gross, F.~Wilczek and A.~Zee,
  Phys.\ Rev.\ D, {\bf 18}: 4684---4692 (1978)

\bibitem{Eichten:1980mw} 
  E.~Eichten and F.~Feinberg,
  Phys.\ Rev.\ D, {\bf 23}: 2724---2744 (1981)

\bibitem{Diakonov:1989un} 
  D.~Diakonov, V.~Y.~Petrov and P.~V.~Pobylitsa,
  Phys.\ Lett.\ B, {\bf 226}: 372---376 (1989)

\bibitem{Diakonov:1983hh} 
  D.~Diakonov and V.~Y.~Petrov,
  Nucl.\ Phys.\ B, {\bf 245}: 259---292 (1984)

\bibitem{Diakonov:1985eg} 
  D.~Diakonov and V.~Y.~Petrov,
  Nucl.\ Phys.\ B, {\bf 272}: 457---489 (1986)

\bibitem{Diakonov:2002fq} 
  D.~Diakonov,
  Prog.\ Part.\ Nucl.\ Phys.,\  {\bf 51}: 173---222 (2003)

\bibitem{Fukushima:1997rc} 
  M.~Fukushima, H.~Suganuma, A.~Tanaka, H.~Toki and S.~Sasaki,
  Nucl.\ Phys.\ Proc.\ Suppl.,\  {\bf 63}: 513---515 (1998)

\bibitem{Chen:1998ct} 
  D.~Chen, R.~C.~Brower, J.~W.~Negele and E.~V.~Shuryak,
  Nucl.\ Phys.\ Proc.\ Suppl.,\  {\bf 73}: 512---514 (1999)

\bibitem{Diakonov:1998rk} 
  D.~Diakonov and V.~Petrov,
  Phys.\ Scripta, {\bf 61}: 536---543 (2000)

\bibitem{Georgi:1990um} 
  H.~Georgi,
  Phys.\ Lett.\ B, {\bf 240}: 447---450 (1990)

\bibitem{Mannel:1991mc} 
  T.~Mannel, W.~Roberts and Z.~Ryzak,
  Nucl.\ Phys.\ B, {\bf 368}: 204---217 (1992)

\bibitem{Foldy:1949wa} 
  L.~L.~Foldy and S.~A.~Wouthuysen,
  Phys.\ Rev.,\  {\bf 78}: 29---36 (1950) 

\bibitem{Korner:1991kf} 
  J.~G.~K\"orner and G.~Thompson,
  Phys.\ Lett.\ B, {\bf 264}: 185---192 (1991)

\bibitem{Pobylitsa:1989uq} 
  P.~V.~Pobylitsa,
  Phys.\ Lett.\ B, {\bf 226}: 387---392 (1989)

\bibitem{Shuryak:1981ff} 
  E.~V.~Shuryak,
  Nucl.\ Phys.\ B, {\bf 203}: 93---115 (1982)

\bibitem{Son:2015bwa} 
  H.~D.~Son, S.~i.~Nam and H.-Ch.~Kim,
  Phys.\ Lett.\ B, {\bf 747}: 460---467 (2015)

\bibitem{PDG2016}
C. Patrignani et al. (Particle Data Group), Chin. Phys. C,
\textbf{40}: 100001 (2016) 

\bibitem{Gray:2005ur} 
  A.~Gray, I.~Allison, C.~T.~H.~Davies, E.~Dalgic, G.~P.~Lepage,
  J.~Shigemitsu and M.~Wingate, 
  Phys.\ Rev.\ D, {\bf 72}:  094507 (2005)

\bibitem{Chiu:2007km} 
  T.~W.~Chiu {\it et al.} [TWQCD Collaboration],
  Phys.\ Lett.\ B, {\bf 651}:  171---176 (2007)

\bibitem{Meinel:2010pv} 
  S.~Meinel,
  Phys.\ Rev.\ D, {\bf 82}: 114502 (2010)

\bibitem{Dowdall:2011wh} 
  R.~J.~Dowdall {\it et al.} [HPQCD Collaboration],
  Phys.\ Rev.\ D, {\bf 85}:  054509 (2012)


\bibitem{Richardson:1978bt} 
  J.~L.~Richardson,
  Phys.\ Lett.\ B, {\bf 82}: 272---274 (1979)

\bibitem{Buchmuller:1980su} 
  W.~Buchmuller and S.~H.~H.~Tye,
  Phys.\ Rev.\ D, {\bf 24}: 132---156 (1981)

\bibitem{Kim:2005jc} 
  H.-Ch.~Kim, M.~M.~Musakhanov and M.~Siddikov,
  Phys.\ Lett.\ B, {\bf 633}: 701---709 (2006)

\bibitem{Goeke:2007bj} 
  K.~Goeke, M.~M.~Musakhanov and M.~Siddikov,
  Phys.\ Rev.\ D, {\bf 76}:  076007 (2007)

\bibitem{Goeke:2007nc} 
  K.~Goeke, H.-Ch.~Kim, M.~M.~Musakhanov and M.~Siddikov,
  Phys.\ Rev.\ D, {\bf 76}:  116007 (2007)

\bibitem{Chu:1994vi} 
  M.~C.~Chu, J.~M.~Grandy, S.~Huang and J.~W.~Negele,
  Phys.\ Rev.\ D, {\bf 49}: 6039---6051 (1994)

\bibitem{Negele:1998ev} 
  J.~W.~Negele,
  Nucl.\ Phys.\ Proc.\ Suppl.,\  {\bf 73}: 92---104 (1999)

\bibitem{DeGrand:2001tm} 
  T.~A.~DeGrand,
  Phys.\ Rev.\ D, {\bf 64}:  094508 (2001)

\bibitem{Faccioli:2003qz} 
  P.~Faccioli and T.~A.~DeGrand,
  Phys.\ Rev.\ Lett.,\  {\bf 91}:  182001 (2003)

\bibitem{Brambilla:2004jw} 
  N.~Brambilla, A.~Pineda, J.~Soto and A.~Vairo,
  Rev.\ Mod.\ Phys.,\  {\bf 77}: 1423---1496 (2005)

\bibitem{'tHooft:1976fv} 
  G.~'t Hooft,
  Phys.\ Rev.\ D, {\bf 14}: 3432---3450 (1976)
  [Phys.\ Rev.\ D, {\bf 18}: 2199(1978)]


 \bibitem{Hiyama2003}
E. Hiyama, Y. Kino and M. Kamimura, Prog. Part. Nucl. Phys.,
\textbf{51}: 223 (2003).
\end{thebibliography}
\end{document}